\documentclass[a4paper,fleqn,usenatbib]{mnras}

\usepackage{amsmath}	
\usepackage{txfonts}

\usepackage[T1]{fontenc}
\usepackage{ae,aecompl}

\usepackage{graphicx}	

\def\simless{\mathbin{\lower 3pt\hbox
     {$\rlap{\raise 5pt\hbox{$\char'074$}}\mathchar"7218$}}}   
\def\simmore{\mathbin{\lower 3pt\hbox
     {$\rlap{\raise 5pt\hbox{$\char'076$}}\mathchar"7218$}}}   

\title[The $\Gamma$ - $t_{\rm lag}$ correlation in BHBs]{The photon-index - time-lag correlation in black-hole X-ray binaries}

\author[P. Reig et al.]{
Pablo Reig$^{1,2},$\thanks{E-mail: pau@physics.uoc.gr}
Nikolaos D. Kylafis$^{2,1}$,
Iossif E. Papadakis$^{2,1}$,
and Mar\'{\i}a Teresa Costado$^{3}$
\\
$^{1}$IESL, Foundation for Reseach and Technology-Hellas, 71110, 
   		Heraklion, Greece\\
$^{2}$Physics Department \& Institute of Theoretical \& Computational Physics, University of Crete, 70013, 
   		Heraklion, Greece\\
$^{3}$Instituto de Astrof\'{\i}sica de Andaluc\'{\i}a, CSIC, Apdo 3004, 18080 Granada, Spain
}

\date{Accepted XXX. Received YYY; in original form ZZZ}

\pubyear{2017}

\begin{document}
\label{firstpage}
\pagerange{\pageref{firstpage}--\pageref{lastpage}}
\maketitle

\begin{abstract}

We have performed a timing and spectral analysis of a set of black-hole binaries
to study the correlation between the photon index and the time lag of the hard
photons with respect to the soft ones. We provide further evidence that the
timing and spectral properties in black-hole  X-ray binaries are coupled.  In
particular, we find that the average time lag increases as the X-ray emission
becomes softer. Although a correlation between the hardness of the X-ray
spectrum and the time (or phase) lag has been  reported in the past for a
handful of systems, our study  confirms that this correlated behaviour is a global
property of black-hole X-ray binaries.  We also demonstrate that the
photon-index - time-lag correlation can be explained as a result of inverse
Comptonization in a jet. 

\end{abstract}




\begin{table*}
\centering
\caption{List of sources. The fifth column gives the final number of
observations used (blue empty circles in Fig.~\ref{hid}) over the total number analysed. 
}
\label{sources}
\begin{tabular}{llccc} 
\hline
Object 		&Outburst	&MJD interval	 &Total exposure  &Observations \\
		&epoch		&		 &time (ks)       &lags/total 	        \\
\hline
4U\,1543--475	&2002		&52443--52565	&275.4		&16/106   \\
\hline
Cyg X--1	&2003-2004	&52693--53182	&367.4		&135/151	\\
\hline
GRO\,J1655--40	&2005		&53423--53685	&2238.3 	&66/501 \\
\hline
GX\,339--4	&2002		&52311--52884	&495.6  	&45/267 	\\
 		&2004		&53050--53498	&565.2  	&139/328	\\
 		&2007		&53769--54678	&572.6  	&177/347 	\\
\hline
H\,1743--322	&2008		&54746--54789	&66.3   	&34/182 	\\
		&2009		&54980--55039	&84.9   	&16/182 	\\
\hline
XTE\,J1720--318	&2003		&52654--52849	&261.9    	&32/99   	\\
\hline
XTE\,J1550--564	&2000		&51644--51741	&128.7   	&25/66  	\\
		&2001--2004	&51938--53163	&306.5   	&64/101 	\\
\hline
XTE\,J1650--500	&2002		&52159--52447	&327.9  	&46/182 	\\
\hline
\hline
\end{tabular}
\end{table*}

\section{Introduction}

The vast majority of black-hole binaries (BHBs) are transient X-ray sources
that spend much of their life in a dormant state. Occasionally, they become
the brightest sources of the X-ray sky. The X-ray outbursts last from weeks
to months and are characterized by a fast rise and a slower decay, although
there is no unique morphological type \citep{chen97}. During these
outbursts, the timing and spectral properties change, defining a series of
states. In a plot of X-ray luminosity (or count rate) as a function of
hardness, the so-called hardness-luminosity (HLD) or hardness-intensity
diagram (HID), the sources trace a characteristic $q$-shaped pattern, where
each state occupies a specific part of the diagram \citep[e.g.][and references
therein]{kylafis15a,kylafis15b}.

The physical processes that give rise to the X-ray emission are thermal and
non-thermal. When the flux is high, the X-ray spectrum is dominated by a thermal
component that is generally modeled as a multi-temperature blackbody with a
characteristic temperature of 0.1--0.5 keV. The origin of this component is
believed to be the innermost part of a geometrically thin, optically thick
accretion disk \citep{shakura73}, which is formed by mass transfer from the
optical companion through the Lagrangian point of the Roche lobe. When the flux
is low, the X-ray spectrum is dominated by a non-termal component, which is best
described as a power law with an exponential cutoff. The origin of this
component is believed to be inverse Compton scattering of low-energy photons off
very energetic electrons.  However, neither the origin of the photons that are
up-scattered nor the geometry and properties of the comptonizing medium are
known with certainty. The Comptonizing medium could be an optically thin, very
hot "corona" in the vicinity of the compact object
\citep{titarchuk80,hua95,zdziarski98}, an advection-dominated accretion flow
\citep{narayan94,esin97}, a low angular momentum accretion flow
\citep{ghosh11,garain12}, the base of a radio jet
\citep{band86,georganopoulos02,markoff05} or a more extended jet
\citep{reig03,giannios05}. The most likely origin for the source of low-energy
photons is the  blackbody photons from the accretion disk. Other possible
scenarios involve cosmic microwave background photons or synchrotron photons
emitted by jet electrons \citep{giannios05,mcnamara09}.

Because the blackbody emission is at low energies and power-law emission at
high energies, the states where these components are dominant are called
soft and hard, respectively. The power-law spectrum observed in the
hard state is commonly referred to as the hard tail. 

Optically thick, compact, steady radio emission is detected during the 
hard state, while optically thin radio flares occur during transitions from
the hard to the soft sate. The association of compact radio jets and 
X-ray hard tails with a specific region of the source in the
hardness-intensity diagram is well documented 
\citep{fender04,migliari06,migliari07,fender09, miller-jones10}. Both the
radio emission and the strength of the hard tail become weaker at higher
accretion rates. Radio and hard X-rays show the strongest intensity in the
hard states of BHBs. For the formation and destruction of jets in
black-hole and neutron-star binaries, the reader is referred to
\citet{kylafis12}.

Spectral information alone cannot reveal the accretion geometry: different
models can  result in very similar energy spectra \citep{nowak11}.  Significant
advances in our understanding of how the X-ray mechanism in BHBs  works can be
achieved when we combine variability and spectral information. The discovery of
correlations between  a) the characteristic frequencies of the noise components
in the power spectra and the photon index of the power-law in the energy
spectra  \citep{dimatteo99,shaposhnikov09,shidatsu14},  b) the broad-band noise
components with luminosity \citep{reig13},  c) the broad-band noise components
with the disk temperature and radius \citep{kalamkar15},  d) the photon index
with the time lag \citep{pottschmidt03,grinberg14},  and the
anticorrelations e) between the fraction of up-scattered photons and the
characteristic frequency of quasi-periodic oscillations (QPO) \citep{stiele13},
and  f) between the cut-off energy and the phase lag
\citep{altamirano15,reig15}  constitute convincing evidence that the timing and
spectral properties of the sources are closely linked.  In particular, the time
lag between two  different energy bands sets tight constraints on the models of
the X-ray production. In most cases, positive lags, i.e. hard photons delayed
with respect to soft photons, are observed. The lag follows a power-law
dependence with Fourier frequency $\tau \propto \nu^{-0.8\pm0.1}$
\citep{crary98,nowak99a,cassatella12}. Positive lag calculated at energies above
$\sim$2 keV has been traditionally attributed to inverse Comptonization. In
order to acquire their energy, harder photons scatter more times than softer
photons, hence  staying longer in the Comptonizing medium before they escape. In
this context, time  lags simply signify the light-travel time of photons within
the Comptonizing region. A different explanation of the time lags was offered by
\citet{kotov01} and \citet{arevalo06}. In their model, the lags result from
viscous propagation of mass accretion fluctuations within the inner regions of
the disk. 


In this work, we focus on the correlation between the shape of the spectral
continuum and the time lag in the hard state by analysing data corresponding to
twelve outbursts of eight BHBs. So far, this correlation has been
reported in detail for  Cyg X-1 \citep{pottschmidt03,bock11,grinberg14} and  GX
339--4 \citep{nowak02,altamirano15}. A decrease of the time lag as the
source spectrum becomes harder has also been observed in XTE\,J1650--500
\citep{kalemci03} and 4U 1543-47 \citep{kalemci05}. \citet{kalemci02} combined
observations from several systems into one diagram, highlighting the fact that
this correlation may be a common feature in BHBs. Our results confirm the
universality of the correlation.

\begin{figure*}
       \includegraphics[width=16cm]{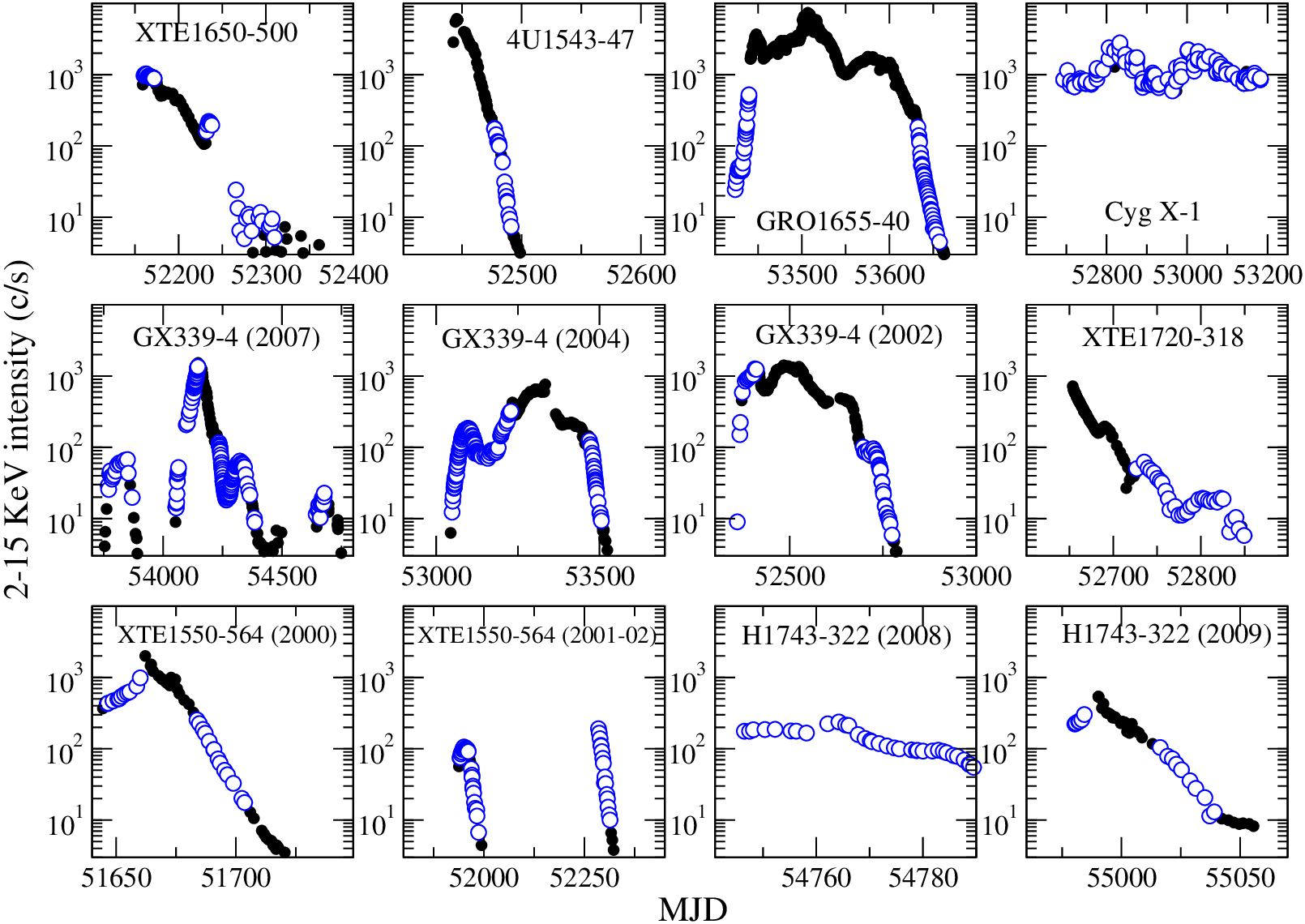}
    \caption{Light curves. Each point corresponds to one
    observation. The blue empty circles identify the observations used
    in the final lag-spectral analysis. }
    \label{outburst}
\end{figure*}
\begin{figure*}
       \includegraphics[width=16cm]{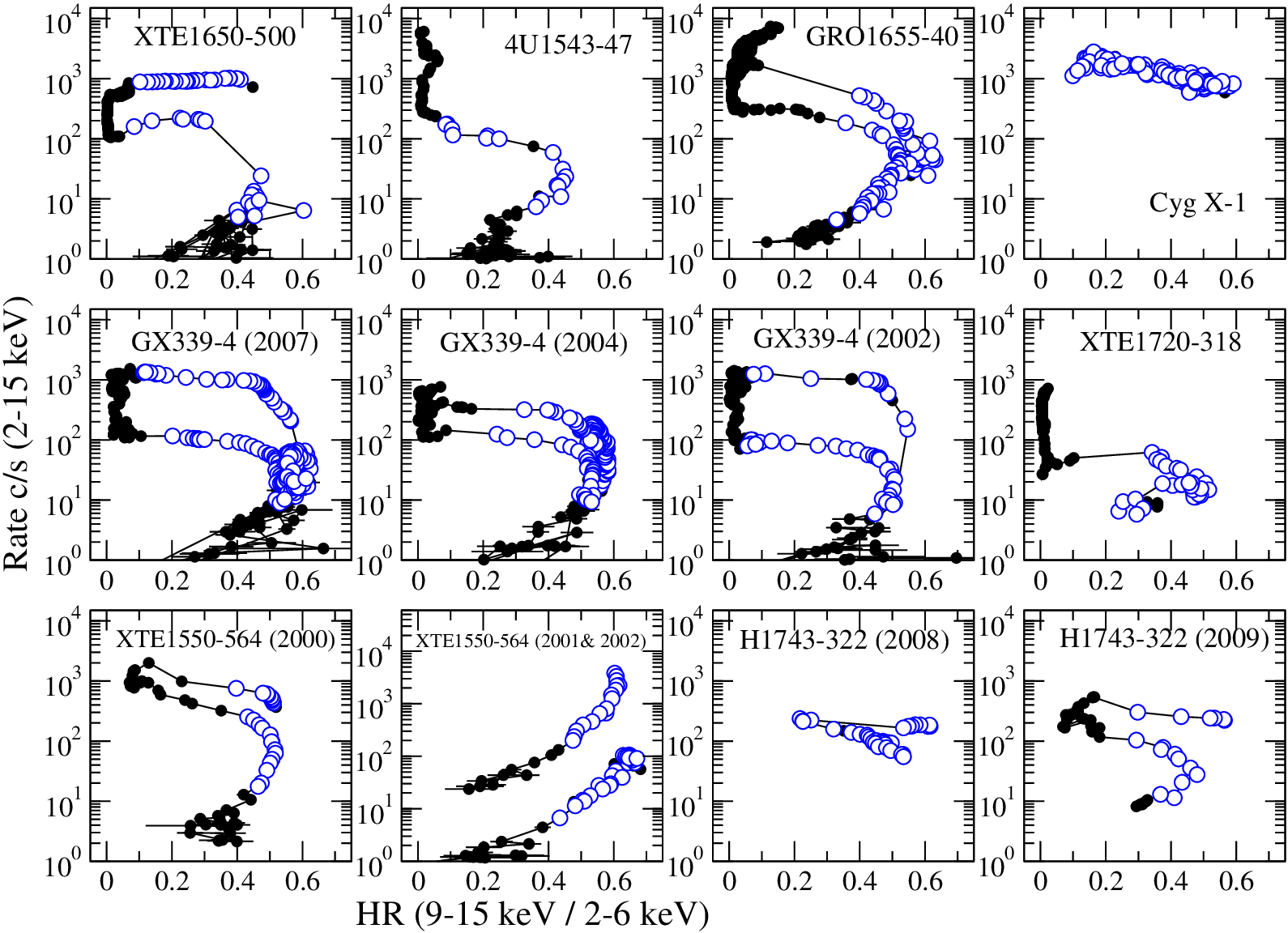}
    \caption{Hardness-intensity diagrams. Each point corresponds to one
    observation. The blue empty circles identify the observations used
    in the final lag-spectral analysis. We show two mini-outbursts of
    XTE\,J1550--564 in the same panel. The 2002 points were shifted up  
    ($\times$ 20) for clarity.}
    \label{hid}
\end{figure*}

\section{Observations and data analysis}

All the observations of the sources analyzed in this work 
(Table~\ref{sources}) are available in the Rossi X-ray Timing Explorer ({\it
RXTE}) archive. All these sources (except for Cyg X--1, { see e.g.  
\citealt{grinberg13}) are X-ray transients, that are only  detected when they
undergo an outburst. During the outburst, the X-ray luminosity increases by
three or four orders of magnitude.   We have also included  Cyg X--1 because it
is the best studied BHB and will allow us to verify our results with previous
work. The sample of sources is not intended to be complete. We selected sources
with well-sampled outbursts.  

The {\it RXTE} mission  was operational in the period 1996--2012 and
carried three detectors: the Proportional Counter  Array (PCA), the
High-energy X-ray Experiment (HEXTE), and the All Sky Monitor  (ASM). The
PCA \citep{jahoda06} covered the energy range 2--60 keV and consisted of
five identical coaligned gas-filled proportional units (PCU), giving a total
collecting area of 6500 cm$^2$ and provided an energy resolution of 18\% at 6
keV. The HEXTE \citep{rothschild98} was constituted of 2 clusters of 4
NaI/CsI scintillation counters each, with a total collecting area of 2 $\times$
800 cm$^2$, sensitive in the 15--250 keV band, with a nominal energy resolution
of 15\% at 60 keV. The ASM \citep{levine96} scanned about 80\% of the sky
every orbit, allowing monitoring on time scales of 90 minutes or longer in
the energy range 1.3--12.1 keV.

Due to {\it RXTE}'s low-Earth orbit, the observations consisted of a number of
contiguous data intervals or pointings interspersed with observational gaps
produced by Earth occultations of the source and passages of the satellite
through the South Atlantic Anomaly. Data taken during satellite slews, passage
through the South Atlantic Anomaly, Earth's occultation, and high voltage
breakdown were filtered out. Typically, an observation consists of a continuous
stretch of data with a  duration of 1000--3000 seconds, although shorter and
longer duration intervals are present in the data.

The timing analysis was performed using data from all the PCUs, except when a
comparison of the intensity levels was relevant (e.g. Fig.~\ref{hid}), in which
case we used PCU2 because it was the  best calibrated PCU and the one that was
in operation all the time. The spectral analysis was performed using data from
the PCU2 and HEXTE. We used Clusters A and B until July 2006, when the
detector's on- and off-source modulation of cluster A began to experience
problems, and Cluster B only from that period until December 2010 when it
suffered the same malfunction.  Owing to these problems, we did not analyse data
after December 2010.

\subsection{Timing analysis}



The timing analysis consisted of the extraction of light curves, which allowed
us to examine the outburst profiles and compute the power spectral density (PSD)
and the time lag.  The light curve was divided into 64-s segments and a
Fast Fourier Transform was computed for each segment. The final power spectrum
is the average of all the power spectra obtained for each segment. The  power
spectra were logarithmically rebinned in frequency and normalized such that the
integral gives the squared rms fractional variability
\citep{belloni90,miyamoto91}. A multicomponent model, consisting of three
Lorentzian functions, fit most of the PSD adequately. When a strong
quasi-periodic oscillation (QPO) was present, then another Lorentzian component
was added. We fitted the PSD for the only purpose to calculate the broad-band
(0.01--30 Hz) rms to select the hard states.

We computed the time lag between  photons in the energy range  9--15 keV (hard
band) with respect to the reference band 2--6 keV (soft band) for each
observation interval.   We generated light curves $x_i(t)$ with time bin size
$2^{-7}$ s for each of these two bands.   Each light curve was divided into 64-s
segments and its Fourier transform $X_i(\nu_j)$ was computed for each segment.
The Fourier transforms were used to compute the average cross-spectrum, defined
as $C(\nu_j)=X_1^*(\nu_j)X_2(\nu_j)$, where the asterisk denotes complex
conjugate. The phase lag between the signals in the two bands at Fourier 
frequency $\nu_j$ is $\phi(\nu_j)=arg[C(\nu_j)]$ [the position angle of $C(\nu_j)$ in
the complex plane]  and the corresponding time lag $t_{\rm lag}(\nu_j)=
\phi(\nu_j)/2\pi\,\nu_j$. We calculated an average cross  vector $<C(\nu_j)>$ by
averaging the complex values over multiple adjacent 64-s  spectra at each
frequency. The final time lag resulted from the average of the time lags in the
frequency range 0.05--5 Hz. 

\begin{figure}
	\includegraphics[width=\columnwidth]{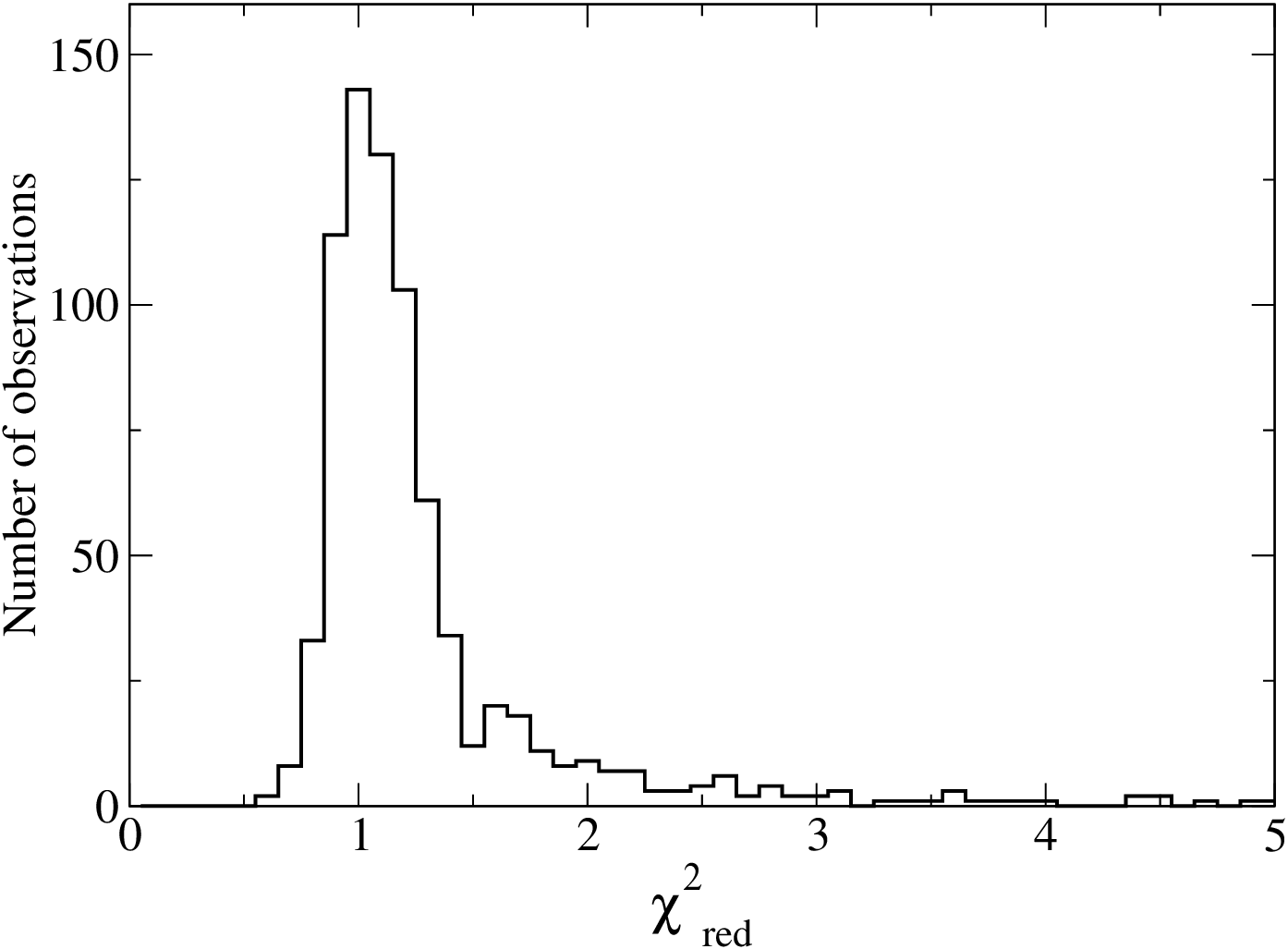}
    \caption{Distribution of the reduced $\chi^2$ after fitting the spectra with
    the broken power law model.}
    \label{chi2}
\end{figure}

\subsection{Selection of the hard state}
\label{selection}

In this work we are interested in the hard and hard-intermediate states of
BHBs, because these are the states when inverse Comptonization dominates the X-ray
emission.  We identified the hard and hard-intermediate states using timing
information only.  By ignoring spectral information, we avoid the introduction
of possible bias in the selection procedure, as our objective is to investigate
the relationship between timing and spectral parameters.

Hard states are characterised by strong variability of the form of band-limited
noise \citep{munoz-darias11,belloni11}.  As representative of the hard  and
hard-intermediate states, we selected observations with an rms above 10\%. The
rms was obtained from the 2--15 keV PSD in the frequency range 0.01--30 Hz. To
have an acceptable signal-to-noise, we discarded observations with less than 20
c\,s$^{-1}$ in the 2--30 keV range. Likewise, we only considered observations
that contain at least ten data segments of 64-s length each.

To better visualize the selected data, we show the outburst light curves in
Fig.~\ref{outburst} and the hardness-intensity diagrams (HID) in Fig.~\ref{hid}.
To create the HID, we extracted light curves with a time resolution of 16 s
using the {\em Standard 2} configuration in three different energy bands. We
used the  counts in the energy ranges 2--6 keV and 9--15 keV to derive the
hardness ratio HR, while the intensity (Y-axis) corresponds to the energy range
2--15 keV. We obtained one point per observation by averaging the count rate  of
the corresponding light curve.   In Figs.~\ref{outburst} and \ref{hid}, we
indicate with circles (filled and empty) all the observations initially
considered (Table~\ref{sources}). With blue empty circles we display the
observations selected for further analysis, that is those observations that
fulfill the selection criteria stated above. As can be seen in Fig.~\ref{hid},
our selection criteria based solely on the timing properties result in
observations that correspond to hard and intermediate states and do not include
observations in the soft state, that is, points in the vertical left branch of
the HID.

XTE\,J1550--564 (and H\,1743--322) sometimes displays mini-outbursts with low
but significant increases in intensity above the quiescent state. In these
mini-outbursts the source remains in the hard state and does not show state
transitions. Two of these mini-outbursts of  XTE\,J1550--564 have been plotted
in the same panel in Fig.~\ref{hid}. In this panel, we shifted up ($\times 20$)
the data points of the 2002 outburst for clarity.

\begin{figure}
	\includegraphics[width=\columnwidth]{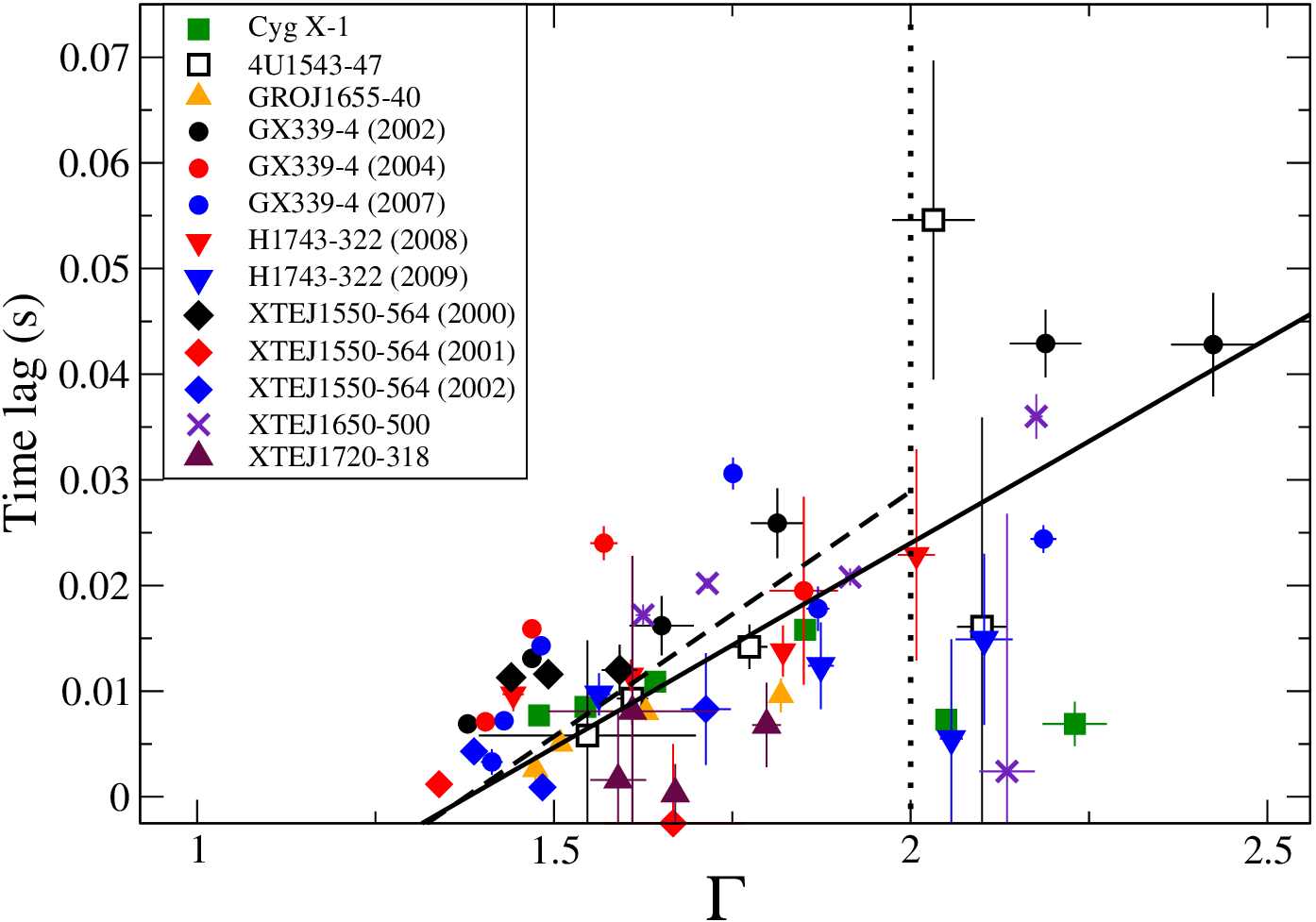}
    \caption{The correlation between the photon index, $\Gamma$, and the
    average time lag, for all observations. Each point is the
    average of the observations within a bin in HR of size 0.1. in The time lag was obtained between the
    9--15 keV photons and 2--6 keV photons and averaged in the frequency
    range 0.05--5 Hz. The solid line is the fit to all data points while the dashed line 
    is the fit to points with $\Gamma < 2$.}
    \label{gamma-lag}
\end{figure}

\begin{figure*}
\begin{tabular}{cc}
	\includegraphics[width=\columnwidth]{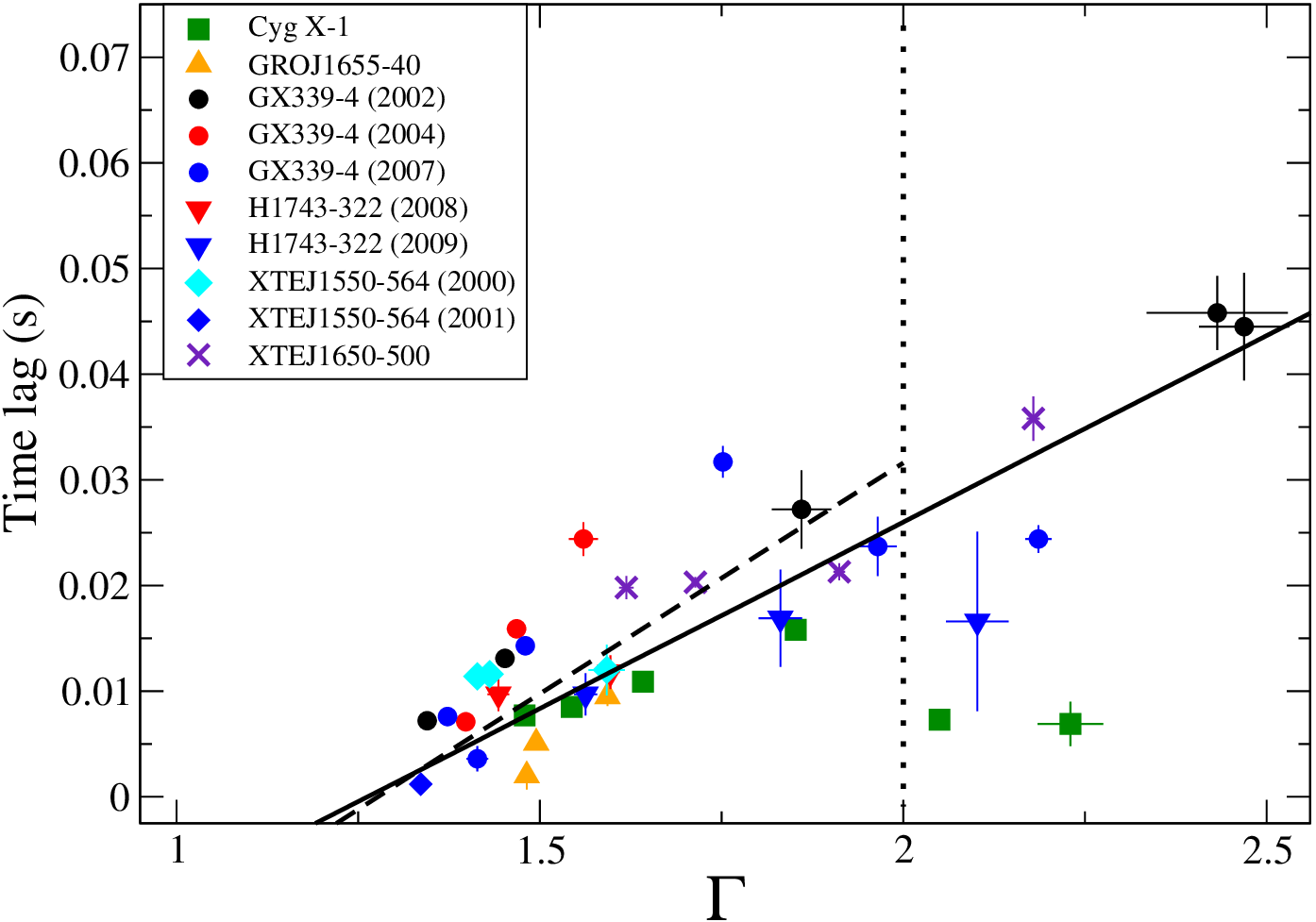} &
	\includegraphics[width=\columnwidth]{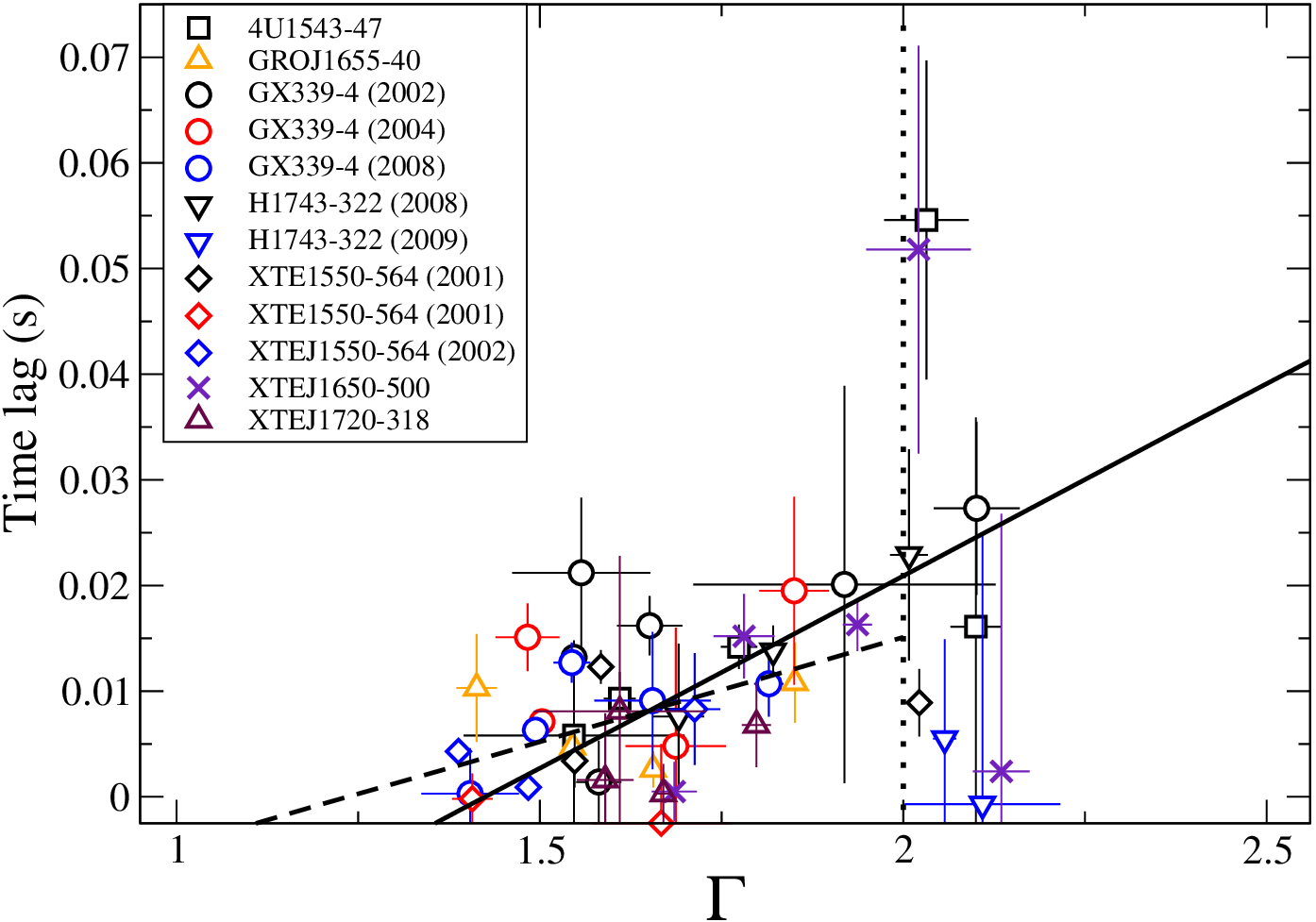}\\
 \end{tabular}
   \caption{The correlation between the photon index, $\Gamma$, and the
    average time lag, considering only observations during the rise (left) 
    and the decay (right) of the outburst. The solid line is the fit to all data points while 
    the dashed line is the fit to points with $\Gamma < 2$. }
    \label{gamma-lag-rise}
\end{figure*}

\subsection{Spectral analysis}
\label{specana}

For the selected observations, we obtained the energy spectra using the standard
modes for both  PCA and HEXTE.  PCA provides 128 channels, while HEXTE provides
64 channels to cover the full energy range. The PCA spectrum becomes background
dominated above $\sim 25$ keV, while the calibration below $\sim 3$ keV is
uncertain. Because the hard state is a low-intensity state, the HEXTE spectra
begin to be dominated by the background at higher energies. Therefore the
overall energy range considered in our spectral analysis was 3--150 keV. 

We wished to characterise the spectral continuum dominated by Comptonization in
as much an independent way as possible, regardless of the underlying physical
process. For this reason, we chose the photon index of a power-law model to be
representative of this component. Thus, we first fitted the spectra with a
simple model that consisted of an absorbed power law and an exponential cutoff.
A narrow  Gaussian component (line width $\sigma \simless 0.9$ keV) was added to
account for the iron emission line at around 6.4 keV.  The hydrogen column
density was fixed to the values provided by \citet{dunn10}. For Cyg X--1 we used
the value given by \citet{hanke09}. This model left significant residuals below
10 keV.  Substantial improvement was achieved with the use of a broken power-law
model. Figure~\ref{chi2} shows the distribution of the reduced $\chi^2$.
The distribution peaks at around 1. The reduced $\chi^2$ of 85\% of the fits
lie in the range $0.7 \leq \chi^2 \leq 1.3$, while 91\% of the observations
have  $0.7 \leq \chi^2 \leq 2$. The broken power-law model has been
successfully used to fit the spectra of Cyg X--1 \citep{wilms06,grinberg14}.  
The parameters of the broken power-law model  are a soft photon index, a hard
photon index, and a break energy. The break energy is typically found at
$\sim10$ keV. We used the best-fit hard photon index in our analysis. In some
sources, the X-ray continuum shows a roll over at high energy, which was modeled
with an exponential cutoff. As expected, no blackbody component was needed to
obtain a good fit.

\begin{table*}
\centering
\caption{Results of the linear regression and correlation analysis for all $\Gamma$ and for 
$\Gamma< 2$. Here $r$ is Pearson's correlation coefficient, 
$N$ is the number of points, and $p$ the probability that the null hypothesis
(no correlation) is true. }
\label{cores}
\begin{tabular}{lccccc|lccccc} 
\hline
&\multicolumn{5}{c}{All points}	&\multicolumn{5}{c}{ Points with $\Gamma < 2$} \\
	&Intercept	&Slope			&$r$    &N	&$p$-value		&Intercept 	 &Slope 	       &$r$    	&N      &$p$-value    \\
	&		&			&    	&	      & 		&  		 &		       &    	&	&		      \\
\hline
All   &$-0.053\pm0.012$ &$0.039\pm0.008$      &0.57	&57	&$4.5\times10^{-6}$     &$-0.064\pm0.016$ &$0.046\pm0.010$	&0.51  	 &45	&$3.6\times10^{-4}$	 \\
Rise  &$-0.045\pm0.004$ &$0.035\pm0.003$      &0.73	&37	&$3.6\times10^{-7}$     &$-0.056\pm0.010$ &$0.044\pm0.007$	&0.72  	 &30	&$7.0\times10^{-6}$	 \\
Decay &$-0.052\pm0.033$ &$0.036\pm0.020$      &0.44	&47	&$1.9\times10^{-3}$    	&$-0.025\pm0.025$ &$0.020\pm0.016$	&0.36 	 &39	&$2.6\times10^{-2}$     \\
\hline
\end{tabular}
\end{table*}

\section{Results}

Figure~\ref{gamma-lag} shows the average time lag as a function of the photon
index. In this figure,  we rebinned the observations according to their position
in the HID (Fig.~\ref{hid}), i.e. we divided the $HR$ range  from 0.0 to 0.7
into bins of size 0.1 and obtained the weighted average time lag and photon
index in the corresponding bin. Thus each point in Fig.~\ref{gamma-lag} is the
average of a varying number of observations. If the average is formed out of two
points, then the error is obtained as $\sigma=(\sigma_1^2+\sigma_2^2)^{1/2}$,
where $\sigma_1$ and $\sigma_2$ are the individual errors of the two points.
When the bin contains only one point, we retained the original error. To study
the posibility of different behaviour during the rise and the decay of the
outbursts, we selected observations corresponding to each one of these two
cases separately, and rebinned them as before. The relationship between the time
lag and the photon index for the rise and decay is shown in
Fig.~\ref{gamma-lag-rise}.

Although these figures exhibit large dispersion, there is a general trend
showing that the time lag increases as the photon index increases.  For $\Gamma
\simmore 2$, that is when the source  leaves the pure hard state and enters
the hard-intermediate state, before crossing the jet line, the relationship
becomes more complex, with some sources showing a reversed trend, i.e. the lags
decrease when the photon index increases.  This behaviour is well documented in
Cyg X--1 \citep{grinberg14}. Not all the sources show a turning point, but this
might be due to the fact that different sources may reach the soft state at
different $\Gamma$. Even different outbursts of the same source may or may not
show the drop in $t_{\rm lag}$ at larger $\Gamma$, as in GX\,339--4
\citep{altamirano15}.

We now investigate whether the observed trend between the photon index and the
time lag is statistically significant by performing a correlation and a linear
regression analysis on the three groups of points considered, namely the whole
sample (Fig.~\ref{gamma-lag}) and the rise and decay of the outburst
(Fig.~\ref{gamma-lag-rise}), separately.  In addition, for each one of these
three groups, we carry out the analysis considering all data points and those
with $\Gamma < 2$. The latter case would correspond to a pure hard state.  The
results are presented in Table~\ref{cores}. First, we study whether the two
variables correlate by computing the Pearson's correlation coefficient, $r$, and
the probabilty of the null hypothesis, $p$-value, i.e. that there is no
correlation between the two variables.  We find a significant positive
correlation ($r=0.57$) when we consider all data points (Fig.~\ref{gamma-lag}).
The correlation becomes stronger ($r=0.73$) when we consider the points during
the rise phase of the outbursts (left panel of Fig.~\ref{gamma-lag-rise}).
During the decay (right panel of Fig.~\ref{gamma-lag-rise}), the correlation is
weaker ($r=0.44$) than in the other two cases. The positive correlation remains
strong when we consider the rise points with $\Gamma < 2$ (left panel in
Fig.~\ref{gamma-lag-rise}), despite the fact that the number of points is
smaller. In contrast, an even weaker  correlation is found for the decay data
set and $\Gamma < 2$.

We also performed a linear fit to the data. There is no apriori reason to choose
one of the two variables as the independent variable. The lags and the photon
index are not directly correlated in the sense that one can be considered the
cause and the other one the effect. The correlation between time lag and
spectral continuum most likely arises because the two variables correlate with
another unknown physical parameter (e.g. optical depth, accretion rate).
Moreover, the intrinsic scatter of the data dominates over the errors arising
from the observations and the measurement process. Under these circumstances,
the preferred linear regression method is the bisector of the two lines that
correspond to the least-square fit of Y on X and X on Y following the
prescription by \citet{akritas96}, which takes into account both the individual
errors and the intrinsic scatter. The linear regression is shown in
Figs.~\ref{gamma-lag} and \ref{gamma-lag-rise} as solid lines (all $\Gamma$) and
dashed lines ($\Gamma < 2$) and the results are presented in Table~\ref{cores}.

When we include points in both the hard and hard-intermediate states, i.e. all
$\Gamma$, the slope of the regression is not consistent with zero, irrespective
of the phase of the outburst, although for the decay data, the slope deviates
from zero at less than $2\sigma$. When we consider points with $\Gamma < 2$, 
the slopes of the data sets that include the entire outburst and that of the
rise are steeper than when all $\Gamma$ are considered. In contrast, that of the
decay group is flatter. However, we notice that the slope of the three groups 
is consistent within errors with the values when all data points are taken into
account. 

The extrapolation of the best-fit values for $\Gamma \simless 1.2$ predicts
negative lags (Figs.~\ref{gamma-lag} and \ref{gamma-lag-rise}), which is at
odds with Comptonization models. In terms of the outburst evolution, such
extrapolation would mean entering the quiescent state. However, studies of the
quiescent state of BHBs show that as the systems fade from the low/hard state
toward quiescence, their X-ray spectra become softer with typical values of
$\Gamma$ in the range 2--3
\citep{tomsick01,tomsick04,kalemci05,wu08,sobolewska11,armas-padilla13}.  Thus,
the correlation must break down at around $\Gamma\sim1.2$. 

Our results can be summarised as follows: i) we detect a significant and
strong positive correlation between time lag and spectral slope during the rise
phase of the outbursts in BHBs. ii) The significance of the correlation is
lower during the decay. This may simply be a consequence of the poorer
statistics as a result of lower count rates. During the decay, the X-ray
intensity is typically an order of magnitude lower than during the rise. 
iii) The correlation does not change significantly  when we consider
observations with $\Gamma < 2$ only, and  iv) the correlation breaks
down when the source enters the quiescent state.

Next, we investigate whether Comptonization in a jet can explain these results.

\section{A jet model}


All BHBs show radio emission that points toward evidence of the presence of
a compact radio jet when they are in their quiescent, hard, and
hard-intermediate states \citep{fender12}. Over the years we have developed
a jet model to explain the spectral and timing properties  of BHBs
\citep{reig03,giannios04,giannios05,kylafis08,reig15}.  

The jet model used here is the same as the one used in \citet{reig15},
which in turn, is based on the model described in \citet{kylafis08},
with the addition of an acceleration zone.  The jet has an acceleration
region of height $z_1$ beyond which the flow has a constant
velocity equal to $v_0$. Hence, the flow velocity in the jet is given by

\begin{equation}
\label{accel-vel}
v_{\parallel}(z) =
\begin{cases}
(z/z_1)^p ~ v_0  & \text{if } \, \, \, 0 < z \leqslant z_1 \\
v_0              & \text{if}  \, \, \, \, \, z > z_1 
\end{cases}
\end{equation}

The radius of the jet at height $z$ is $R(z) = R_0 (z/z_0)^{1/2}$, which
corresponds to a parabolic jet. The density outside the acceleration region
is $n_e(z)=n_0(z_0/z)$. Within the acceleration region, the density is
obtained through the continuity equation $n_{\rm acc}(z)=(z_1/z)^p \,\,
n_e(z)$.

The parameters of the model are:  the optical depth along the jet's axis
$\tau_{\parallel}$, the width of the jet at its base $R_0$,  the parallel, 
$v_0$, and perpendicular, $v_{\perp}$, components of the velocity, or
equivalently the Lorentz factor $\gamma=1/\sqrt{1-(v_0^2+v_{\perp}^2)/c^2}$,
the distance $z_0$ of the bottom of the jet
from the black hole, the total height $H$ of the jet, the size $z_1$ and
exponent $p$ of the acceleration zone, and the temperature  $T_{bb}$ of the
soft-photon input. Our assumption of a characteristic Lorentz $\gamma$ in
the flow is equivalent to considering the smallest Lorentz factor of a steep
power-law distribution \citep{giannios04,giannios05}.

Physically, a variation in optical depth is equivalent to a change in the
density at the base of the jet. Hence, $\tau_{\parallel}$  is the prime
parameter that drives the changes in the photon index. A change in $R_0$
corresponds to a change in the size of the jet, and time lags are strongly
affected by changes in $R_0$. Finally, the variation in $v_{\perp}$ (or
$\gamma$) mimics the process of Compton cooling, which is related to the cutoff
at high energy. 

The top panel of Fig.~\ref{gamma-lag-model} shows the $\Gamma$--$t_{\rm
lag}$ correlation for different models.  The empty red circles correspond to 
pairs of the model parameters $\tau_{\parallel}$ and $R_0$, with constant
$\gamma=2.24$. This value of $\gamma$ was chosen because it matches the cutoff
energy of $\sim100$ keV, observed in GX\,339--4  \citep{motta09}. The parameters
varied in the ranges  $2 \leqslant \tau_{\parallel} \leqslant 11$ and $40
\leqslant R_0/r_g \leqslant 900$, where $r_g=GM/c^2$ is the gravitational radius
of the black hole. The models (empty red circles) were selected to match the
best-fit line to the data (left panel in Fig.~\ref{gamma-lag-rise}).  Figure
\ref{gamma-lag-model}, bottom panel, displays the data shown in the left panel
of Fig.~\ref{gamma-lag-rise} and the solid line of Fig.~\ref{gamma-lag-model},
top panel., which in this case, is the best-fit line to the models. 
Similarly, we computed  pairs $(\Gamma,~ t_{\rm lag})$ keeping $R_0$ constant at
two different values: $R_0 = 100 r_g$ (empty blue squares) and $R_0 = 300 r_g$
(filled green squares) in Fig.~\ref{gamma-lag-model}, top panel. In these two
cases, we varied the model parameters $2 \leqslant \tau_{\parallel} \leqslant
11$ and  $2.0 \leqslant \gamma \leqslant 2.8$. The dotted line is the best
linear fit to the $R_0 = 100 r_g$ models and the dashed line to the $R_0 = 300
r_g$ ones.   In all the models, the rest of the parameters are  fixed at the
following reference values:  $z_0 = 5 r_g$, $H=10^5 r_g$, $v_0=0.8 c$, $z_1= 50
r_g$, $p=0.5$, and $T_{bb} = 0.2$ keV.


We can reproduce the $\Gamma$ - $t_{\rm lag}$ correlation by  changing the width
of the jet at its base and the optical depth along the jet's axis.   In
contrast, we cannot reproduce the correct slope with a fixed width at the base
of the jet. Changing the optical depth and the perpendicular component of the
jet velocity, without changing the width of the jet can reproduce the range of
photon index, but not the range of time lag. 

We also find that to explain the $\Gamma$ - $t_{\rm lag}$  correlation,
$\tau_{\parallel}$ and $R_0$ should vary in a highly correlated way  following a
power-law function, $\tau_{\parallel} \propto R_0^{\alpha}$ with index
$\alpha=-0.57$ (Fig.~\ref{tau-width}). This  index is smaller than the
one found in  \citet{kylafis08} for Cyg X--1. The difference can be attributed to
the acceleration region, which was absent in \citet{kylafis08}.  The
acceleration zone was introduced to explain the correlation between the cut-off
energy and phase lags in GX\,339--4 \citep{reig15}. As shown in \citet{reig15},
for a given value of $\tau_{\parallel}$, the models with the acceleration zone
produce shorter lags than when acceleration occurs instantaneously. The natural
way to increase the lags is by increasing the size of the jet.



\begin{figure}
	\includegraphics[width=\columnwidth]{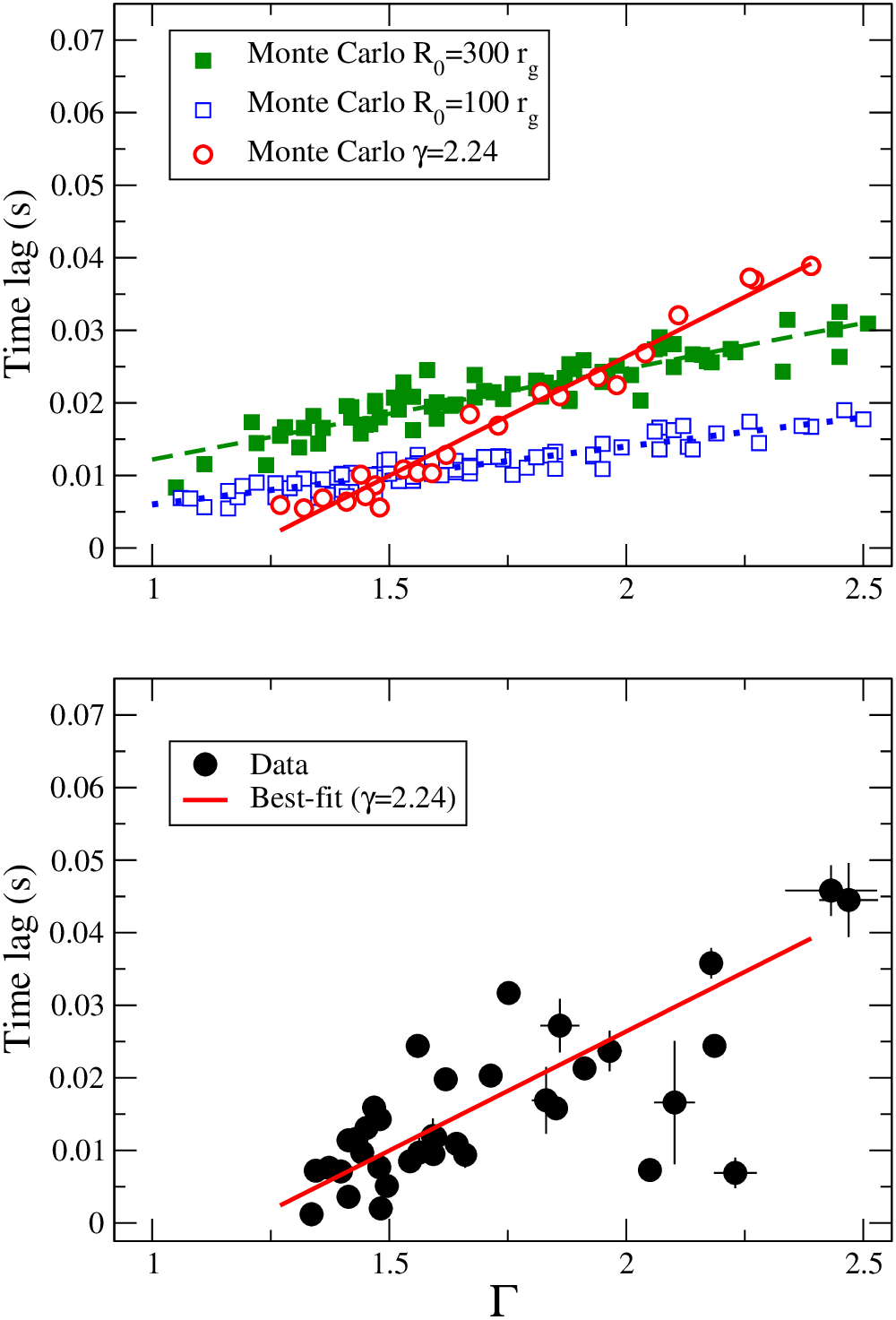}
    \caption{The correlation between the photon index, $\Gamma$, and the
    average time lag. In the upper panel, the points correspond to different
    jet models obtained from Monte Carlo simulations with a fixed value of 
    the Lorentz factor $\gamma$ and variable width at the base of the jet
    $R_0$ (circles) and fixed $R_0$ and variable $\gamma$ (squares). 
    The lines represents the best linear fit to the points. In the bottom panel, 
    the observational data (circles) from the left panel of Fig.~\ref{gamma-lag-rise} 
    and the best-fit from the Monte Carlo simulation
    (solid line in the upper panel) are plotted together.}
    \label{gamma-lag-model}
\end{figure}


\section{Discussion}

Extensive studies of Cyg X--1 \citep{nowak02,pottschmidt03,bock11,grinberg14}
and GX\,339--4 \citep{altamirano15} have shown that the time lag  increases
with decreasing hardness, that is, as the source transits from the hard to the
intermediate state. A similar behaviour has been reported for
XTE\,J1650-500 \citep{kalemci03} and 4U\,1543-47 \citep{kalemci05}. Our study
reveals that this correlated behaviour  is a global property of BHBs. The
$\Gamma$ - $t_{\rm lag}$ correlation imposes stringent constraints on all the
models that seek to explain the behaviour of BHBs. A model that can successfully
explain the time delay between two energy bands, should also predict the right
spectrum.

In Comptonization models, the lags are related to the average number of
scatterings of the low-energy photons with highly energetic electrons. Hence,
the time lag is related to a characteristic length scale of the medium which
upscatters the observed photons.  \citet{nowak02} modeled the spectrum of GX
339--4 with a hybrid coronal model and found that higher coronal compactness,
defined as the energy released by the Comptonizing medium divided by its radius
\citep[see][for a review]{coppi99}, is associated with shorter time lags. The
higher the compactness parameter, the harder the spectra. In principle,
compactness changes are achieved solely by varying the coronal radius. A large
corona would be associated with high compactness, i.e., harder spectra, but
also, following the geometric interpretation, with larger lags, contrary to what
is observed. \citet{nowak02} suggested a jet-like model, where the corona is
simply the base of the jet.  The correlation between compactness and lags in GX
339--4 was observed when the source went from the soft to the hard state.  This
transition would be associated with the formation of a jet.
Time lags could be generated via propagation along the length of the jet.

Alternatively, \citet{kalemci02} explained the correlation by invoking a hotter
medium (``corona") in the hard state, and a cooler one as the source transits to
the soft state. In a very hot medium, the time lag becomes insignificant
because the number of scatterings is roughly the same independent of the energy.
The problem with a uniform static corona is the lack of a heating mechanism to
maintain a highly energetic electron population away (hundreds of gravitational
radii) from the black hole as required by the magnitude of the observed lags
($\sim0.1$ s).

In this work we demonstrate that Comptonization in the jet can reproduce
satisfactorily the observed relationship between the X-ray spectral continuum
emission and the time lag of hard photons with respect to softer ones.  In fact,
the correlation between the photon index and the time lag suggests that the lag
originates in the jet, that is, the same region as where the hard X-rays are
formed. The identification of the Comptonizing medium is a matter of intense
debate with corona, ADAF, and jet as the prime candidates. The fact that the
lags disappear or decrease abruptly when the source moves to the soft state,
i.e., when the radio emission weakens, strongly supports a jet model. 

To understand how Comptonization in the jet can explain the  time-lag -
photon-index correlation, we need to understand the  dependence of the lags and
the photon index on the main parameters of the jet, $\tau_{\parallel}$, $R_0$,
and $\gamma$. At high optical depth, the disk photons penetrate the base of the
jet superficially. Because of the high density, the mean free path is short and
the photons sample a small region at the base of the jet before they escape. The
photon suffers a large number of scatterings on short length scales. Thus
the time lag between hard and softer photons is small and the spectrum is hard
(small $\Gamma$). As the optical depth (or equivalently the density) decreases,
the mean free path increases, and the photons sample a larger volume of the jet.
Thus, the time lag increases and the spectrum becomes softer (larger $\Gamma$).
For the same reasons, as $R_0$ increases while keeping $\tau_{\parallel}$
constant, the lags increase and the photon index decreases \citep{reig03}, i.e.
the photons sample a larger volume and more scatterings take place in a larger
jet. The scatter in the $\Gamma$ - $t_{\rm lag}$ correlation   could then be
related  to differences in the jet's characteristics such as optical depth,
size, and/or velocity. 

The X-ray spectral continuum of some sources displays an exponential cutoff in
the hard state. The value $E_{\rm cut}$ varies from source to source in the
range $\sim$50--200 keV. In our jet model, the cutoff energy strongly depends on
the parameter $\gamma$ (Lorentz factor), i.e. the perpendicular component of the
jet velocity, because it determines the maximum energy gain by the soft photons
\citep{giannios04}. In this work we find that the  $\Gamma$ - $t_{\rm lag}$
correlation can be reproduced  with different fixed values of $\gamma$. All the
sets of models displayed in Fig.~\ref{tau-width} reproduce the $\Gamma$ -
$t_{\rm lag}$ correlation satisfactorily. If we plotted the corresponding
lines in the bottom panel of Fig.~\ref{gamma-lag-model}, they would differ
by less than ~1-2 line widths. Figure~\ref{tau-width}  shows that for a given
$\tau_{\parallel}$, wider jets are required at lower velocities. Interestingly,
the slope $\alpha$ of the relation between $\tau_{\parallel}$ and $R_0$ remains
the same, irrespective of the value of $\gamma$. We conclude that the
differences observed in the cutoff energy can be explained by invoking differet
jet velocity. We refer the reader to \citet{reig15} for a study of the
dependence of the cutoff energy with $\gamma$.

In previous work, we have shown that Comptonization of low-energy disk photons by
energetic electrons in a jet can reproduce many of the timing and spectral
properties of BHBs in the hard state. By changing a small number of the
parameters of the jet such as the density, size or velocity, we have been able
to quantitatively explain: i) the emerging spectrum from radio to hard X-rays
\citep{giannios05}; ii) the evolution of the photon index and the time
(phase)-lags as a function of Fourier frequency \citep{reig03}; iii)  the
narrowing of the autocorrelation function with photon energy \citep{giannios04}
iv) the correlation observed in Cyg X-1 between the photon index and the average
time lag \citep{kylafis08}; v) the correlation between the cutoff energy of the
power law and the phase lag of hard photons with respect to soft ones in GX
339--4 \citep{reig15}. In this work,  we have shown that the  correlation
between the photon index and the average time lag is not exclusive of Cyg X--1,
but appears in most, if not all, BHBs and have demonstrated that Comptonization
in a jet can reproduce it.   We speculate that the ultimate physical mechanism
that triggers the  change in the jet parameters is the Cosmic Battery
\citep{contopoulos98,kylafis12}, which creates the poloidal magnetic field
needed for the formation of the jet. In any case, inverse Comptonization appears
as the only mechanism capable to explain all the above results  in a
self-consistent way. 

\begin{figure}
	\includegraphics[width=\columnwidth]{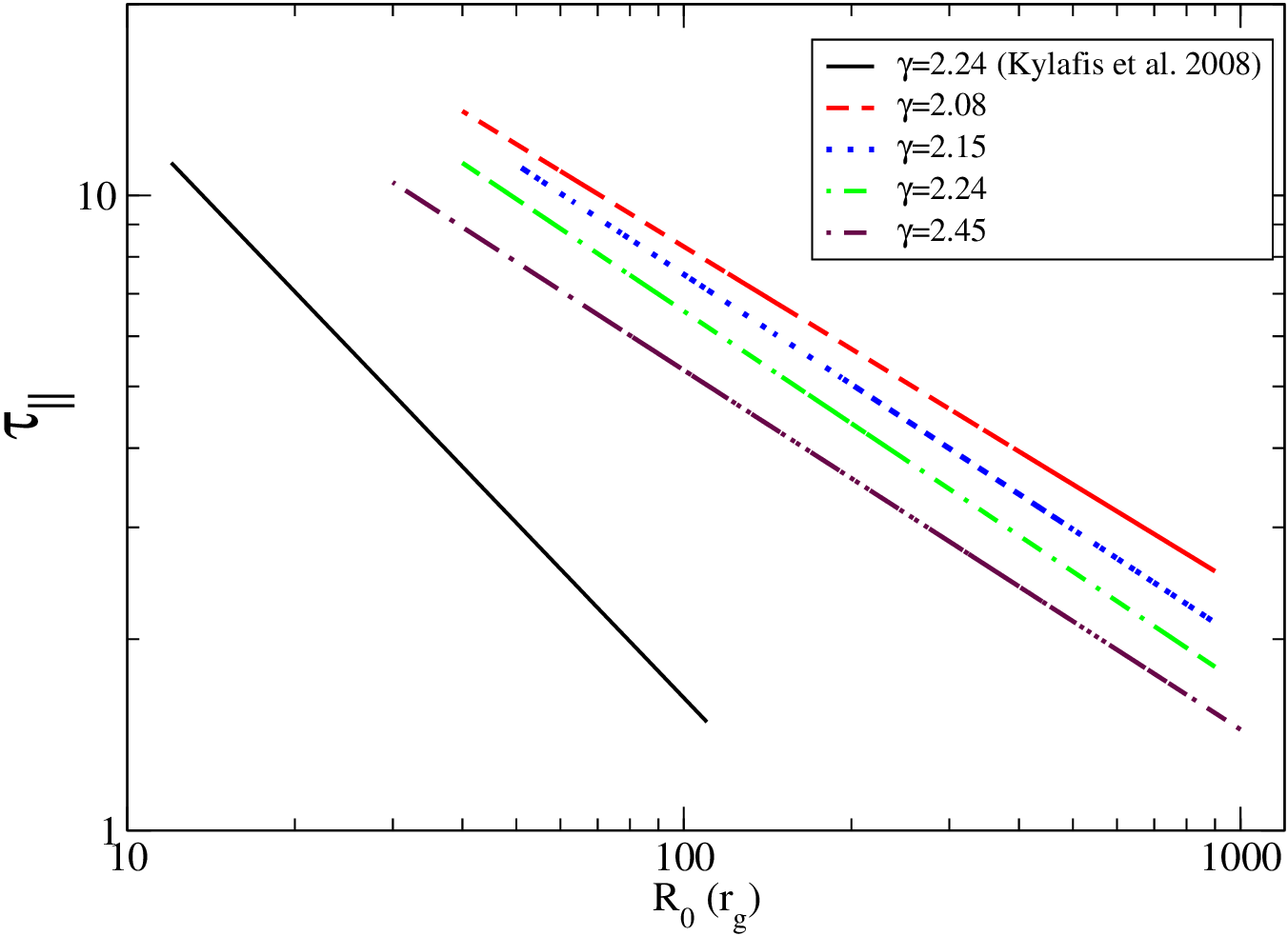}
    \caption{Relationship between the optical depth, $\tau_{\parallel}$, and the width at 
    the base of the jet, $R_0$, for the models that reproduce the observations. Four
    different sets of models (discontinuous lines) are shown. Each set (line) was 
    created by varying $\tau_{\parallel}$ and $R_0$, while $\gamma$ remained fixed 
    at the value indicated. For comparison, we also show (continuous line) the relation 
    between optical depth and jet width in the absence of an acceleration region 
    from \citep[from][]{kylafis08}. }
    \label{tau-width}
\end{figure}

\section{Conclusions}

We have performed an X-ray timing and spectral analysis of twelve outbursts of
eight black-hole binaries and found that the slope of the hard spectral
continuum  correlates with the time lag of hard photons with respect to softer
photons.  As the source evolves in the hard state the spectral continuum becomes
softer and the time lag increases. The correlation is particularly strong and
significant during the rise of the outburst. Remarkably this correlation appears
to be a universal property of BHBs. We demonstrate that Comptonization of
low-energy photons by very energetic electrons in a jet can explain these
results. The different cutoff energy is explained by invoking different jet
velocities.

\section*{Acknowledgements}

NDK acknowledges a useful discussion with Emrah Kalemci regarding the
correlation of time lag with spectral index in his PhD Thesis.  This discussion
sparked the research reported in the present paper. This work was supported in
part by the ``AGNQUEST" project, which is implemented under the ``Aristeia II"
Action of the ``Education and Lifelong Learning" operational programme of the
GSRT, Greece.




\bibliographystyle{aa}
\bibliography{./bhb} 

\bsp	
\label{lastpage}
\end{document}